\documentclass[twocolumn,showpacs,preprintnumbers]{revtex4}
\usepackage[dvips]{graphicx}     
\usepackage{dcolumn}     
\usepackage{bm}

\begin{document}           

\title{Weak ferromagnetism in cobalt oxalate crystals}       
\author{E Romero$^1$, M E Mendoza$^1$ and R Escudero$^2$}
\address{$^1$Instituto de F\'{\i}sica, Universidad Aut\'{o}noma de Puebla, Apartado Postal J-48, Puebla 72570, M\'{e}xico}
\address{$^2$Instituto de Investigaciones en Materiales, Universidad Nacional Aut\'{o}noma de M\'{e}xico, Apartado Postal 70-360,  M\'{e}xico DF 04510, M\'{e}xico} 

\email[Author to whom correspondence should be addressed. RE, email address:] {eromero@sirio.ifuap.buap.mx}

\date{\today }  

\begin{abstract} 
Microcrystals of diaquocobalt(II) oxalate have been synthesized\- by the coprecipitation reaction of aqueous solutions of Cobalt (II) bromide and oxalic acid. Chemical\- analysis and thermal experiments revealed that there is only one phase present.  X-ray powder diffraction studies show  that  this  compound  is  orthorhombic  with space group $Cccm$. Molar susceptibility versus tempe\-rature measurements show the existence of an antiferromagnetic ordering, however, the hysteresis measured in magnetization measurements as a function of magnetic field reveals a weak ferromagnetic behavior.

\end{abstract}      

\maketitle 

\section{Introduction}                
Oxalates dihydrates containing diva\-lent 3d transition elements, with general formula $MC_{2}O_{4}\cdot nH_{2}O$ where M is a metallic ion 2+ 
\cite{Decurtins99,Miller,Ovcharenko,West,Blundell,Simizu-87,Ovan,Awaga00,Hurst04,Keene-04,Price00}, exhibit quasi-one-dimensional magnetic behavior characterized by a dominant magnetic correlation along oxalate-metal ion chains. Kurmoo \cite{Kurmoo-09} has reported a weak antiferromagnetic interaction in these compounds. However, it is expected that in systems like the aforementioned oxalates, with many close chains \cite{Bacsa}, the intra- and inter-chains magnetic interac\-tions cant the spins, distorting the antiferromagnetic order and  producing also a weak ferromagnetism (WF). At low temperatures the WF behavior will be presumably more pronounced.      

Cobalt oxalate dihydrate is a quasi-one dimensional magnetic compound useful to search the existence of WF at low temperature because, it crystallizes in two allotropic forms \cite{Deyrieux}: $\alpha - monoclinic$, space group $C2/c$, $a$ = 11.775 $\mathring{A}$,  $b$ = 5.416 $\mathring{A}$, $c$ = 9.859 $\mathring{A}$ and 
$\beta = 127.9$ $^{\circ }$, and $\beta - orthorhombic$, $Cccm$, 
$a$ = 11.877 $\mathring{A}$, $b$ = 5.419 $\mathring{A}$, 
and $c$ = 15.624 $\mathring{A}$. Both structures are formed by identical infinite chains of $[CoC_{2}O_{4}\cdot 2H_{2}O]$ units, the difference between them lies in the relative displacement of adjacent chains along $b$-axis \cite{Dubernat,Molinier,Deyrieux69,Druet-99}. Each $Co^{2+}$ ion is surrounded by distorted oxygen octahedron, where four oxygens belong to the oxalate anions and the other two to the water molecules. The more stable $\alpha$ structure is the most stu\-died. Its magnetic structure \cite{Sledzinska88a}, has been described as two interpenetrating antiferromagnetic sublattices belonging to the Shubnikov group $P_{b}c$, with parameters $a$, 2$b$, $c$. There was found that below the N\'{e}el temperature $T_{N}=6.1$ $K$, three-dimensional antiferromagnetic long-range order exists\-, corresponding to a collinear magnetic structure with $k=[0,1/2,0]$ and magnetic moments aligned parallel to the $a$ axis.  

In this paper, we will report results on the synthesis, crystal structure, and magnetic meassurement of the ortho\-rhombic $\beta -CoC_{2}O_{4}\cdot 2H_{2}O$ microcrystals. It was determined by M-H measurements from  2 - 200 K, that this phase exhibits WF ordering at T $<$ 7.5 K.

\section{Experimental Methods}  
The synthesis of ortho\-rhombic $\beta$ $-$ cobalt oxalate dihydrate was carried out by coprecipitation reaction of aqueous solutions of cobalt (II) bromide 0.1 M (Aldrich, $99\%$) and oxalic acid 0.00625 M (Aldrich, $\geq99\%$), according\- to the chemical equation,
\begin{equation}
H_{2}C_{2}O_{4(aq)}+CoBr_{2(aq)}\rightarrow CoC_{2}O_{4}\text{\textperiodcentered}2H_{2}O_{(s)}+2HBr_{(aq)}
\end{equation} 

The precipitates were filtered and dried at room temperature.
 
Morphological analyses were performed with a scan\-ning electron microscope (SEM) Cambrige-Leica Stereos\-can 440. Powder X-ray diffraction patterns were acqui\-red using a Siemens D5000 diffractometer, opera\-ting in the Bragg-Brentano
geometry with  $\lambda $(Cu-K$\alpha $) = $1.541$ $\mathring{
A}$, and $2\theta$ scan = 10 - 70$^{\circ }$, with a step size of 0.02$^{\circ }$.   
Thermogravimetric (TG) and differential thermal analy\-sis (DTA) curves were obtained in a SDT-TA Ins\-truments model 2960, in air atmosphere with  heating rate of 5 $^{\circ }$C/min from room temperature up to 600 $^{\circ }$C. Heat capacity was measured between 2 and 300 K in a Quantum Design PPMS system. 
Magnetization measurements were performed in  a Quantum Design MPMS SQUID
magnetometer, MPMS-5. Zero Field Cooling (ZFC) and Field Cooling (FC) cycles were
performed at magnetic intensities of 1T, in the range from  2  to 300 K. Isothermal
magnetization measurements M(H) were obtained at 2, 50, 100, 160 and 200 K. The diamagnetic contribution calculated from Pascal's
constants \cite{Bain08} was $\chi_{Di}=-72 \cdot 10^{-6}cm^{3}/mol$.

\section{Results and Discussion}    

\subsection{Synthesis and Morphology} 

\begin{figure}[btp, h]
\begin{center}
\includegraphics[scale=0.3]{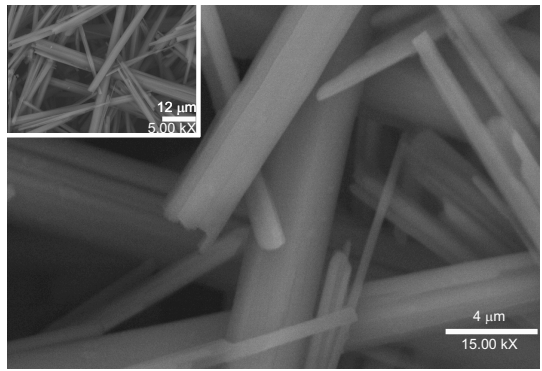}
\end{center}
\caption{(Color online) SEM micrograph of cobalt oxalate needles-like crystallites forming tubular microcrystals, for sample $S4$.} 
\label{Fig1.eps} 
\end{figure}

Five polycrystalline samples were obtained using different reaction times (see Table \ref{tab1}).
Pink cobalt oxalate dihydrate microcrystals were obtained after dry precipitates at room temperature. Chemical analysis by digestion using Inductively Coupled Plasma-Optical Emission Spectrometry (ICP-OEP), by combustion using Thermal Conductivity-Infrared (TCD/IR), and Pyrolysis-IR, give the composition in weight of 35 \% in cobalt, 13.21 \% in carbon, 2.45 \% in hydrogen, and 54.92 \% in oxygen. The morphology shown by all the precipitates is tubular-like (inset in Fig. \ref{Fig1.eps}), generated by self-assembly of needle-like microcrystals as shown in Fig. \ref{Fig1.eps}.  

The observed microtubes have an average length size (L) and
a diameter (D) in the range from  16 - 146.2 and 0.6 - 3.8 $\mu m$
respectively. This morphology is not the usual one for the cobalt oxalates prepared by precipitation methods from homogeneous solutions without additives.           

Some reports concerning to the morphology of 3$d$ ions oxalates prepared by precipitation reactions have been published, Pujol et al. \cite{Pujol} have found that cobalt oxa\-late dihydrate crystals present a rod-like morphology. They used as reac\-tant solutions cobalt nitrate and sodium oxala\-te. Jongen et al. \cite{Jongen00} have also obtained copper oxa\-late with controlled morphology from cushion to rod-like crystals, \-using solutions of copper nitrate and sodium oxa\-late, adding a polymer at  different concentrations. These authors show that the mechanism of growth is by self-assembly, guided by the cooper oxalate crystal structure. Additionaly, nanorods of nickel oxalate were synthesized using  solutions of nickel nitrate and ammonium oxalate adding a cationic surfactant (CTAB) \cite{Vaidya08}. Negative surface charge on the nanorods was observed, it has a bearing on the growth of the rods along the cross-section, especially with surfactant molecules ha\-ving posi\-tively charged headgroups (CTAB). 

It is known that in crystal growth in aqueous solutions, there is a correlation between the ionic strength of the solution\- and the kinetics of crystal faces \cite{Sugi-80}. In order to explain the morphology of our orthorhombic $\beta -CoC_{2}O_{4}\cdot 2H_{2}O$ crystals obtained in this work, and compared with the reported by Pujol, we calcu\-lated the ionic strength of the solutions in both cases using the Debye equation I = $1/2$ $\sum$ $C_{i}$ $Z_{i}^{2}$, where $I$ is the ionic strength, $C_{i}$ and $Z_{i}$ are the concentration and charge of ions $i$, respectively \cite{Ayres70}. Whereas Pujol et al. \cite{Pujol} used cobalt nitrate 0.0052 M and sodium oxalate 0.005 M, in our work we used cobalt bromine 0.00625 M, and oxalic acid  0.1 M. The calculated ionic strength were  0.0306 M \cite{Pujol} and 0.3187 M, (this work) with a difference between them  about one order of magnitude. 

Additionally, in our work the pH at the end of reaction was 1.5, then this high concentration of $H^{+1}$ ions would assembly on the negative surface of the cobalt oxalate crystals growing, and the high ionic strength of the solutions screening the self-assembly of crystallites, giving as a final result the tubular-like morphologies.    

\begin{table}[t,h]  
\begin{center}
\begin{tabular}{|c|c|c|c|c|c|}
\hline
Sample & TG & D($\mu m$) & L($\mu m$) \\ 
\hline
S1 & 12 min & 0.6 & 16.0 \\ 
\hline
S2 & 12 h & 1.4 & 53.3 \\
\hline
S3 & 24 h & 1.5 & 55.4 \\
\hline
S4 & 40 h & 2.1 & 79.9 \\
\hline
S5 & 7 days & 3.8 & 146.2 \\
\hline
\end{tabular}
\caption{Polycrystalline samples obtained using different reaction times. TG denotes growth time, D and L are average diameter and length of the particles, respectively.} 
\label{tab1}
\end{center} 
\end{table}

\subsection{Diffraction pattern} 

Monoclinic $\alpha$ phase of $CoC_{2}O_{4}\cdot 2H_{2}O$ is characterized by the presence of a doublet peak in the XRD powder diagram, it is at $2\theta$ = 18.745$^{\circ }$ \cite{Druet-99}.  

\begin{figure}[btp, h]  
\begin{center}
\includegraphics[scale=0.4]{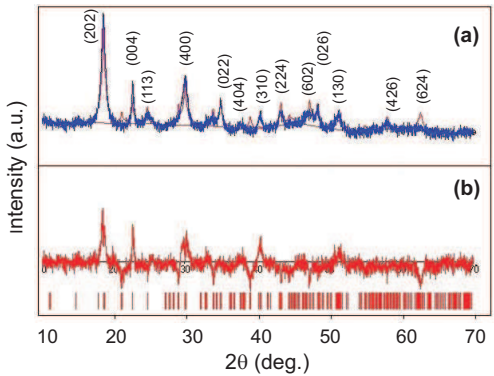}
\end{center}
\caption{(Color online) (a) X-Ray powder diffraction data of sample $S4$ (blue pattern), Miller indices for orthorhombic $Cccm$ (JCPDS file: 25-0250). The theoretic structural (red pattern) was performer using the Win-Rietveld sofware. (b) Refinement diffe\-rence.}
\label{Fig2.eps} 
\end{figure}

Fig. \ref{Fig2.eps}a displays the XRD powder diffraction
pattern of the sample $S4$ (blue pattern); in good agreement with the reported pattern (red lines) for the orthorhombic $\beta $-phase of cobalt oxalate dihydrate (JCPDS file: 25-0250). In all other samples ($S1$, $S2$, $S3$ and $S5$) the same phase was identified.   

\begin{figure}[btp, h]
\begin{center} 
\includegraphics[scale=0.4]{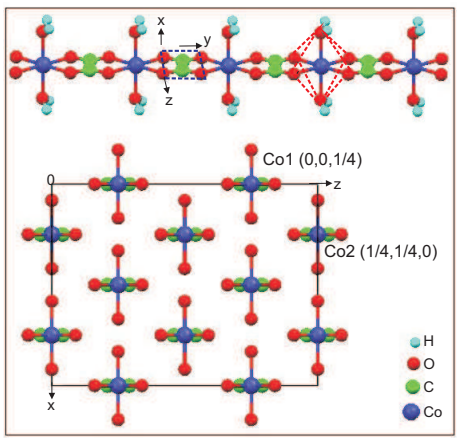} 
\end{center}
\caption{(Color online) Co-oxalate unit cell of the orthorhombic $\beta $-phase of $JCPDS$ file 25-0250, space
group $Cccm$. There are eight cobalt atoms located in two non-equivalent positions designed as Co1 and Co2.} 
\label{Fig3.eps} 
\end{figure}

Rietveld refinement \cite{Cusker99} of the pattern was performed using the Win-Rietveld software, the background was estimated by linear interpolation, and the peak shape was modeled by a Pseudo-Voigt function. Unit cell parameters used were $a$ = 11.877$\mathring{A}$,  $b$ = 5.419$\mathring{A}$, and $c$ = 15.624$\mathring{A}$ \cite{Deyrieux}. The atomic positions  were the reported by Deyrieux et al. for iron-oxalate \cite{Deyrieux69}: $4Co$ $(\frac{1}{4},\frac{1}{4},0),$ $
4Co(0,0,\frac{1}{4}),$ $8C_{oxalate}$ $(\frac{1}{4},\frac{3}{4},0.041),$ $
8C_{oxalate}$ $(0,\frac{1}{2},0.291),$ $16O_{oxalate}$ $(\frac{1}{4}
,0.941,0.089),$ $16O_{oxalate}$ $(0,0.691,0.339),$ $8O_{water}$ $(0.419,
\frac{1}{4},0),$ $8O_{water}$ $(0.169,$ $0,\frac{1}{4})$. The weighted profile and expected residual factors obtained by the Rietveld refinement were R$_{wp}$ = $20.04$ and R$_{exp}$ = $9.95$. The refined
cell parameters determined for the Co-oxalate orthorhombic phase in all samples were $a$ = 11.879(4) $\mathring{A}$, $b$ = 5.421(2) $\mathring{A}$, and $c$ = 15.615(6) $\mathring{A}$. A schematic representation of the unit cell is shown in Fig. \ref{Fig3.eps}. In this figure there are two non-equivalent positions for the cobalt ions, designed as Co1 and Co2, each cobalt ion is shifted res\-pect to other by a translation vector (${\frac{1}{2},\frac{1}{2},0}$).

Finally, mean crystallite dimension $(D)$ for the sample was  calculated with the Scherrer equation (Eq.~\ref{scherrer}) using the $(202)$ peak.   
\begin{equation} 
\label{scherrer}
D=\frac{0.9\lambda }{(B-b)cos\theta }  \label{Scherrer ecuation},
\end{equation} 
where $\lambda $ is the wavelength of the radiation, $B$ is the width of the diffraction line at half intensity maximum, and $b$ is the instrumental broadening, 0.05, $\theta $ is the diffraction angle. The calculated values were 21.4 nm for $S1$, $S2$, and $S3$ samples, 22.7 nm for $S4$, and 26.4 nm for $S5$ sample.

\subsection{Thermal analysis}

\begin{figure}[btp, h]  
\begin{center}
\includegraphics[scale=0.8]{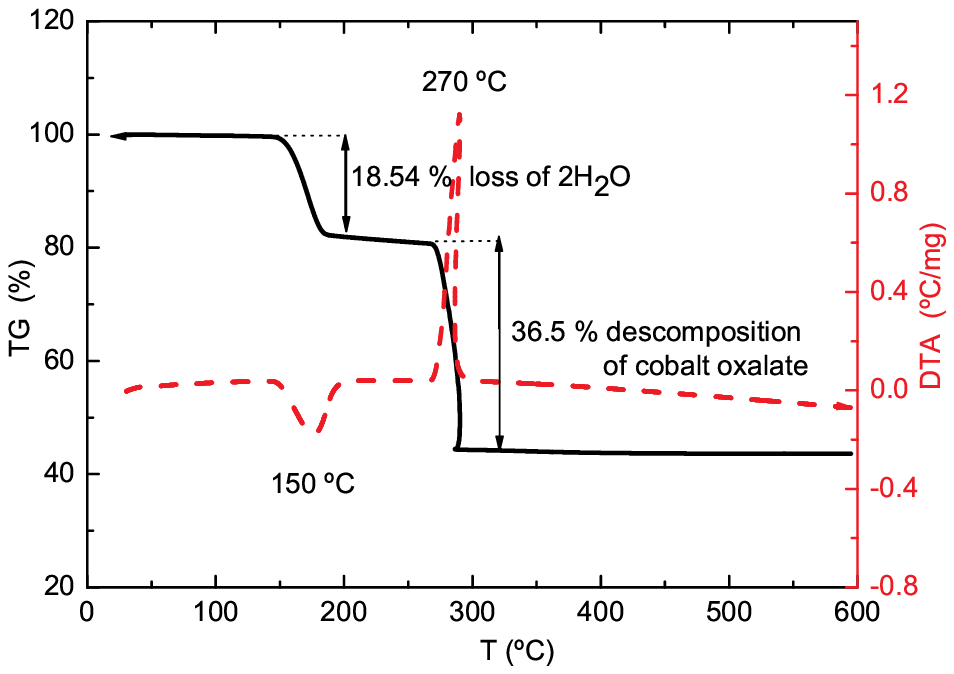}
\end{center}
\caption{(Color online) TG(solid line)/DTA(dash line) curves for $CoC_{2}O_{4}\cdot 2H_{2}O$ (sample $S4$) at 5 $^{\circ }$C/min heating rate.}
\label{Fig4.eps}
\end{figure}

TG studies of all samples of $CoC_{2}O_{4}\cdot 2H_{2}O$ have shown two steps of weight loss, the first one ocurring at about 150 $^{\circ }$C, and the second one at about 270 $^{\circ }$C. Fig. \ref{Fig4.eps} presents the TG curve for sample $S4$. In the first step the weight loss of 18.5 \% corresponds to two water molecules, this is according to the theoretical value of 19.7 \%. The DTA curve associated with this process shows an endothermic peak at about 145 $^{\circ }$C. The dehydration reaction is
\begin{equation}
CoC_{2}O_{4}\cdot 2H_{2}O_{(s)} \rightarrow CoC_{2}O_{4(s)}+2H_{2}O.
\end{equation}

The weight loss of 36.4 \% in the second step may be attri\-buted to the decomposition of the anhydrous cobalt oxa\-late to obtain cobalt oxide. This agrees with the theore\-tical value \cite{Wang05}. The corresponding DTA curve shows an exothermic peak, about $T$ = 267 $^{\circ }$C. The decomposition reaction in this step can be written as:  
\begin{equation}
3CoC_{2}O_{4(s)}+2O_{2} \rightarrow Co_{3}O_{4(s)}+6CO_{2(g)}.
\end{equation}

It is important to observe  that the absence of any additional peak on TG and DTA curves, indicates the high purity of the cobalt oxalate samples.

\subsection{Heat capacity} 

\begin{figure}[btp, h]  
\begin{center}
\includegraphics[scale=0.8]{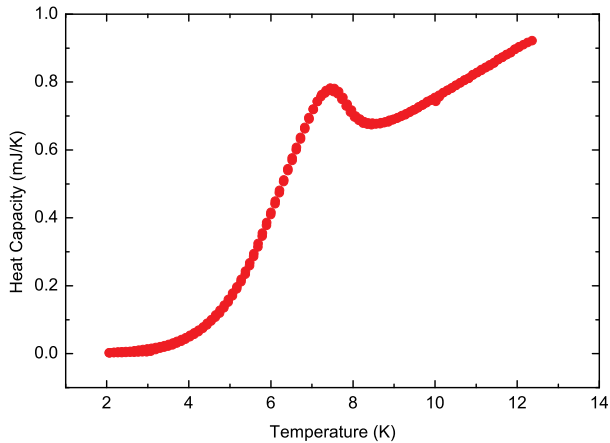}
\end{center}
\caption{(Color online) Heat capacity of $\beta -CoC_{2}O_{4}\cdot 2H_{2}O$, sample $S4$. Peak at $T=7.5 $ $K$. This lambda peak represents the AF transition with onset at about 8.2 K, maximum at 7.5 K.}
\label{Fig5.eps}
\end{figure}

Heat capacity of $\beta-CoC_{2}O_{4}\cdot 2H_{2}O$ (sample $S4$) is plotted as function of temperature in Fig. \ref{Fig5.eps}. The most important feature is the sharp $\lambda $-peak ocurring at about 7.5 $K$. It is important to mention that for  $\alpha-CoC_{2}O_{4}\cdot 2H_{2}O$, the $\lambda $-peak was reported at about $T=6.23$ $K$, indicating the onset of long-range magnetic order \cite{Lukin-95} and consistent with the N\'{e}el temperature $T_{N} = 6.1 + 0.1$ $K$ found in deuterated cobalt oxalate \cite{Sledzinska88a}. So for $\beta-CoC_{2}O_{4}\cdot 2H_{2}O$ we report a N\'{e}el temperature at about 7.5 $K$.  

\subsection{Magnetic measurements} 

\begin{figure}[btp, h]
\begin{center} 
\includegraphics[scale=0.8]{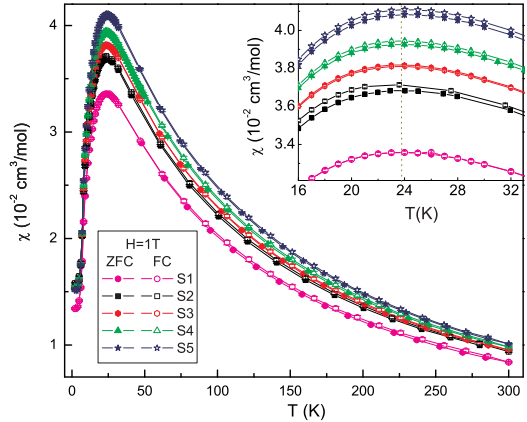}  
\end{center}
\caption{(Color online) Molar
 susceptibility ($\chi(T)$) of 5 samples of $\beta -CoC_{2}O_{4}\cdot 2H_{2}O$, with  maximum
at 24 K, measured for ZFC and FC modes, with H = 1 T. Full
colors (solid symbols) represent ZFC measurements, whereas open symbols is FC mode.}
\label{Fig6.eps}
\end{figure}

The molar susceptibility - Temperature ($\chi(T)$), for the five samples ($S1$-$S5$) is shown in Fig. \ref{Fig6.eps}. It was measured at magnetic field of 1T in both ZFC, and FC modes. All samples have almost the same behavior, at low tempera\-ture the magnetic susceptibility values are small, it increases rapidly as the tempe\-rature increases up to  a maxi\-mum close to 24 K, see inset of Fig. \ref{Fig6.eps}. As the temperature is raised $\chi(T)$ smoothly decreases up to a mini\-mum  at room temperature. Although $\chi_{max}$ values are dif\-ferent in all samples (varying in the range of $3.3$x$10^{-2}$ to $4.08$x$10^{-2}$ $cm^{3}/mol$), they are similar to those reported for $\alpha-CoC_{2}O_{4}\cdot 2H_{2}O$ where $\chi_{max}$ = $3.6$x$10^{-2}$ $cm^{3}/mol$ \cite{Lukin-95}. From room temperature to 100 K a Curie-Weiss behavior clearly can be fitted.  

The maximum on susceptibility at $T_{max}$ = 24 K, is a typical behavior of the susceptibility occurring in low dimensional antiferromagnets \cite{Kurmoo-09,Bonner-64}, this curve is co\-rrectly predicted by 
\begin{equation} 
\label{Xmax}
\frac{\chi_{max}|J|}{g^{2}\mu_{B}^{2}}=0.07346,
\end{equation}

at
\begin{equation} 
\label{Tmax}
\frac{k_{B}T_{max}}{|J|}=1.282,
\end{equation}
with g = 2 (for 3$d$ ions), $\mu_{B}$ = 9.27x$10^{-24}$ and  $k_{B}$ = 1.38x$10^{-23}$, it was obtained a value $|J|$ / $k_{B}$ = 18.72 K, i.e. $|J|$  = 1.6 meV.   

\begin{figure}[btp, h]  
\begin{center}
\includegraphics[scale=0.8]{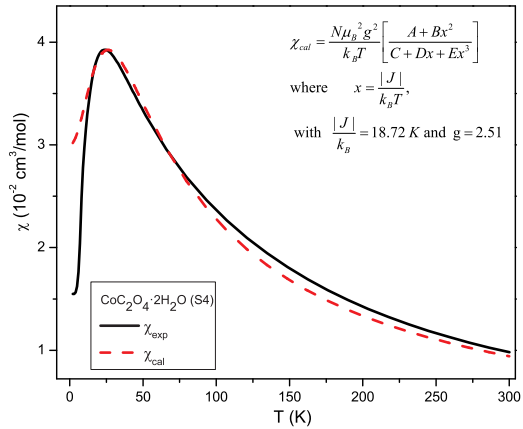}
\end{center}
\caption{(Color online) Molar
 susceptibility ($\chi_{exp}$(T), solid line) of $\beta -CoC_{2}O_{4}\cdot 2H_{2}O$ (sample $S4$) and the fit ($\chi_{cal}$(T), dash line) for antiferromagnetically coupled chain of quantum spins.} 
\label{Fig7.eps} 
\end{figure}

In the case of nickel oxalate chains, Keene et al. \cite{Simizu-87} have modeled the behavior of an antiferromagnetically coupled chain of quantum spins by a polynomial appro\-ximation (Eq.~\ref{Eq.polinomy}).
\begin{equation}  
\label{Eq.polinomy}
\chi_{cal}=\frac{N\mu_{B}^2g^2}{k_{B}T}[\frac{A+Bx^2}{C+Dx+Ex^3}],
\end{equation} where $x$ = $|J|$ / $k_{B}$T. Using this model for the molar susceptibility of $\beta -$cobalt oxalate $\chi_{exp}$ (solid line), in Fig. \ref{Fig7.eps} a good fit (dash line) is obtained with $|J|$ / $k_{B}$ = 18.72 $K$, T (2 $K$ - 300 $K$), g = 2.51,  A =  1.3667, B = 1.36558, C = 1, D = 2.3018 and E = 5.7448. 

The maximum shown in the susceptibility in figures \ref{Fig6.eps} and \ref{Fig7.eps} is a characteristic feature of the effect of spins fluc\-tuations as described by the Heisenberg model \cite{Blundell}. The inset of Fig. \ref{Fig6.eps} shows the molar susceptibility versus tempe\-rature from 16 to 32 K. Here, FC and ZFC modes present a small irreversibility that gives a hysteretic behavior at about 30 K with the maximum at about 24 K. 
This irreversibility corroborates the spin fluctuating characteristics at this temperature. At high temperatures the system tends clearly to be in a paramagnetic state.  
When the temperature decreases thermal energy will be small and spin
fluctuations tend to be canceled and aligned in a minimum energy configuration. 

Our results are similar to those reported for oxalate-cobalt (II) complexes \cite{Ovan,Glerup,Garcia}. However our experimental observations show that this broad maximum may be due to the competition of two different magnetic ordering, an uncompensated antiferromagnetism, and consequently a weak ferromagnetism. 

\begin{figure}[btp, h]  
\begin{center}
\includegraphics[scale=0.8]{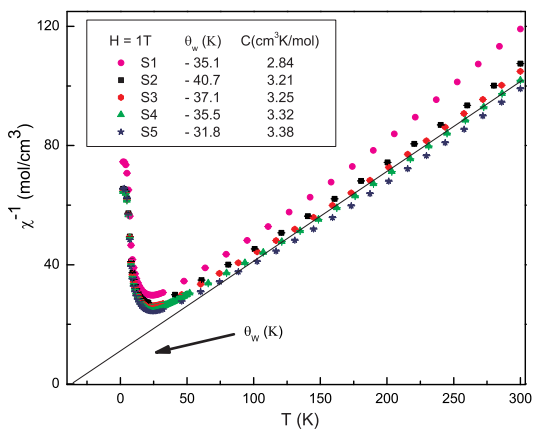}
\end{center}
\caption{(Color online) Inverse susceptibility $\chi ^{-1}$(T) corres\-ponding to 5 samples. Weiss 
temperature $\theta_{\omega}$, changes from -35.8 to -40.7 K. Curie constant also changes from 2.84 to 3.38 $cm^{3}mol^{-1}$K.}
\label{Fig8.eps} 
\end{figure}

In Fig. \ref{Fig8.eps} it is show a plot of the inverse of susceptibi\-lity, $\chi ^{-1}$ as a function of temperature 
at H = 1T, for all samples. 

The analysis of these measurements was performed by fitting a Curie-Weiss behavior above 100 K, where the fit parameters; Weiss temperature $\theta_{\omega}$, and Curie constant C, varying from -31.8 to -40.7 K and from 2.84 to 3.38 $cm^{3}mol^{-1}$, respectively.  

\begin{figure}[btp, h]  
\begin{center}  
\includegraphics[scale=0.8]{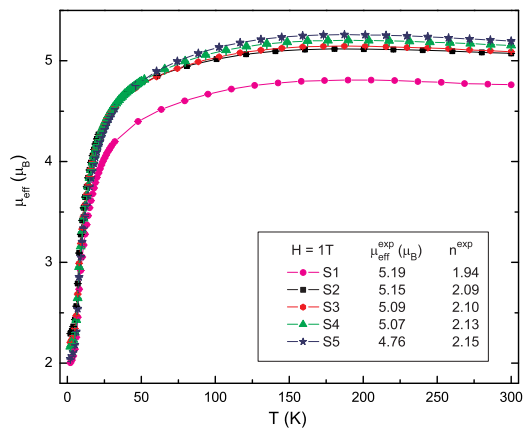} 
\end{center}
\caption{(Color online) Effective Bohr magnetons, $\mu _{eff}$ per mole as a function of temperature. At room temperature the 
va\-lues are between 4.76 to 5.15 $\mu _{B}$. $n$ is the number of unpaired electrons.}
\label{Fig9.eps}  
\end{figure}

These data permit calculate  the effective magnetic moment $\mu _{eff}$ per mole, according to the equation $\mu _{eff}=2.84[C]^{1/2}=2.84[\chi(T-\theta_{\omega}]^{1/2}$. In Fig. \ref{Fig9.eps}, the effective number of Bohr magnetons at room temperature for each sample ($S1$ - $S4$) is  different (4.76, 5.07, 5.09, 5.15, 5.19 $\mu _{B}$). However, it is in agreement with reported values in studies of layered transition metal oxalates \cite{Price03} where for octahedrally coor\-dinated cobalt(II) with a $^{4}T_{1g}$ ground term the observed moment is typically 4.7 - 5.2 $\mu_{B}$. After calculation of $\mu _{eff}$, the number of unpaired electrons $n$ in $\beta-CoC_{2}O_{4}\cdot 2H_{2}O$, using $\mu _{eff}$ = $g[n(n+1)]^{1/2}$ can be calculated. Those va\-lues for each sample are in the range of 1.94 to 2.15.

The decrease in $\mu _{eff}(T)$ on cooling and the large nega\-tive Weiss temperature, indicate an antiferromagnetic coupling between neighboring ions, for all samples ($S1$ - $S5$).     

\begin{figure}[btp, h]  
\begin{center}  
\includegraphics[scale=0.8]{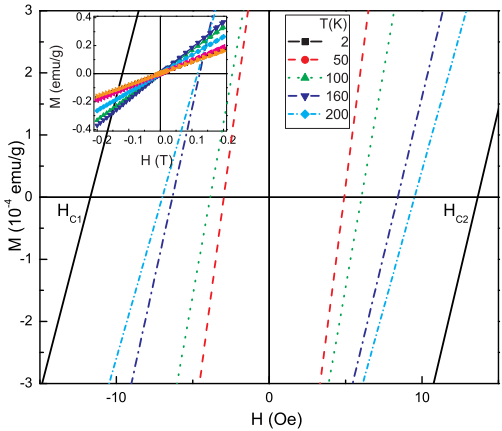} 
\end{center}
\caption{(Color online) Isothermal magnetization measurements M-H from 2 to 200 K shown in the main panel. Inset shows the magnetization at high field 
and different tempera\-tures from 2 to 200 K.  At low fields (main panel) the hysteretic effect is clearly observed. The asymmetric behavior in the coercive field might be related to a exchange bias effect due to an uncompensated or canted antifferromagnetism, but other type of interaction are not discarded as driven by Dzyaloshinsky-Moriya type exchange (DM).}
\label{Fig10.eps}  
\end{figure}

It is important to mention that short-range order is obser\-ved in materials with a low-dimensional magnetic character, where strong magnetic interactions between the nearest ions are along the chains \cite{Price03}. To understand more about the magnetic characteristics of this oxalate, we stu\-died the M-H isothermal measurements in the temperature range from 2 - 200 K, the resulting data is shown in Fig. \ref{Fig10.eps} for the sample $S4$.   

As can be noted in the inset of Fig. \ref{Fig10.eps}, the magnetization never reaches a
saturation value, the M-H curve behaves almost in linear form, thus characteristic of  an AF order. However, a careful observation at low fields, show a small, but measurable hysteretic behavior. 

\begin{figure}[btp, h]   
\begin{center}  
\includegraphics[scale=0.8]{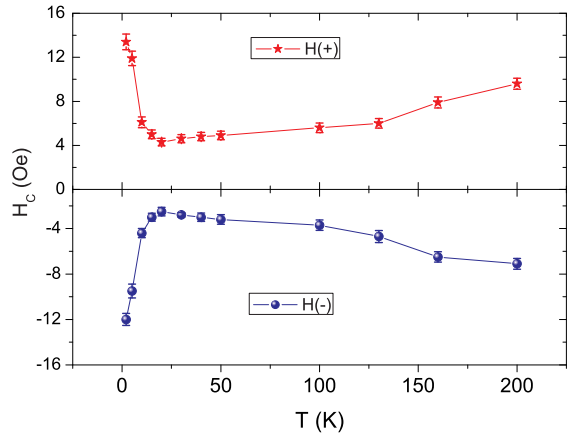}
\end{center}
\caption{(Color online) Coercive Field
vs Temperature as determined by isothermal magnetic measurements for sample $S4$. Note the minimum at about 20 K.} 
\label{Fig11.eps}
\end{figure}

In order to clarify the existence of the hysteretic effect, studies of the coercive field $H_{C}$ as a function of tempera\-ture were performed. Fig. \ref{Fig11.eps}, shows $H_{C}$ versus T for sample $S4$. Here we see the  evidence of the asymmetric beha\-vior related of the coercive field, thus implying an exchange bias effect $H_{C1}$$\neq$
$H_{C2}$ dues to two competing magnetic ordering: AF and a WF. All studied samples have similar behavior. 
This exchange bias can be explained as the effect of inter and intra chains interaction in this compound. The effect of interaction of  metallic ions between chains is canting the spins. This small but measurable exchange bias indicates that in place to have a pure antifferromagnetic order the magnetic order will be distorted by uncompensated antiferromagnetism due to canted spins. Thus  two magnetic orders will be the result: a canted antiferromagnetism, and a weak ferromagnetism. In this study all our studied samples presented similar  behavior. It is important to mention that great care was taken when measuring the exchange bias. Our SQUID magnetometer is provided with a Mu metal shielding. At the moment of performing the magnetization measurements a flux gate magnetometer was used to demagnetize the superconducting coil. This procedure reduces the magnetic field to a very small value to about 0.001 G, or less and the Mu shielding eliminates external magnetic influences, as the earth magnetic field.

In figure \ref{Fig11.eps} we observe that at 2 K the coercive field is about +13 Oe and -13 Oe and decreases rapidly to a mini\-mum value of 3 Oe, at temperaures close to  20 K. Above 20 K the coercive field increases in a smooth form reaching a value of 8 Oe at room temperature. This small but measu\-rable coercive field clearly indicates a system with canted spins, thus a competition between possible DM interaction antiferromagnetism and weak ferromagnetism.  

Evidently the situation at low temperatures is more complicated. Analysis of the susceptibility data for cobalt(II) complexes may show deviation from simple Curie-Weiss behavior \cite{Garcia} by different mechanism: single-ion, orbital moments, spin-orbit coupling, distortions from regular stereochemistry and crystal field and also DM interactions. All these processes clearly may  affect the magnetic properties.   

The ground state of the free Co(II) ion is $^4F$, but the orbital degeneracy is removed in an octahedral crystal field giving a $^4T_{1g}$ ground state. The combined effect of the spin-orbit coupling and the axial or rhombic distortions of the crystal field most often gives rise to six Kramers doublets, two are much lower in energy. When the temperature is low enough, only the ground Kramers doublet is thermally populated and the Co(II) ion can be formally treated with an effective spin S = 1/2 and a very anisotropic g tensor \cite{Lukin-95}. In the case of $\beta -CoC_{2}O_{4}\cdot 2H_{2}O$ this temperature is low enough and is close to 7.5 $K$.

Reports on magnetic studies of some chain and laye\-red six-coordinated cobalt(II) compound \cite{Simizu-87,Garcia} show that, at low temperature, these magnetic systems behave as a collec\-tion of Ising chain with S = 1/2 effective spin coupled ferromagnetically, and/or  with antiferromagneti\-cally interac\-tions \cite{Garcia}. It is necessary to note that the Ising model is restricted to the temperature range where only the ground Kramers doublet are thermally populated, thus (T$<$ 40 $K$). However, at high temperatures  a crossover to the Heisenberg type behavior is expected.

\section{Conclusions}       

Microcrystals of cobalt (II) oxalate were prepared by soft solution chemistry. XRD powder diffraction patterns show a orthorhombic phase. Chemical analysis, DTA and TG studies revealed that the microcrystals have high purity. 
$\chi$-T measurements reveal  the existence of an AF ordering,  by  interaction of coupled chains via inter and intra interactions, and/or DM type-exchange. Those effects of interchain interaction affects the AF coupling, distorting it and canting spins producing a weak ferro\-magnetism ordering. This WF and canted AF is evi\-dent by hysteresis measurements in M-H studies. 

\begin{acknowledgments}
Partial support for this work is gratefully acknowledge to CONACyT, project No.44296/A-1 and Scholarship CONACyT, register No. 188436 for E. Romero; VIEP-BUAP, project No. MEAM-EXC10-G. We also thank F. Morales for providing the Heat Capacity meassurements.
\end{acknowledgments}

\end{document}